\documentstyle[aps,prl,multicol,epsf,psfig]{revtex}
\begin{document}
\draft
\tightenlines
\title{\bf{ Nonlocality in kinetic roughening}}
\author{Sutapa Mukherji\cite{eml1}} 
\address{ Institut f\"ur Theoretische Physik,
Universitat zu K\"oln, D-50937 K\"oln, Germany}
\author{ Somendra M. Bhattacharjee\cite{eml1}} 
\address{Institute of Physics, Bhubaneswar 751 005, India}
\date{\today}
\maketitle
\widetext
\begin{abstract}
We propose a phenomenological equation to describe kinetic roughening
of a growing surface in presence of long range interactions.
The roughness of the evolving surface depends on the long range feature,
and several distinct scenarios of phase transitions are
possible. Experimental implications are discussed.
\end{abstract}
\begin{multicols}{2}

``Suppose that we take a bin and gently and uniformly pour in granular
material.  As the material in the bin builds up we can identify a
surface and ask the question,`What is the magnitude of the fluctuation
in the height of surface (measured from the base of the bin)?' Also of
interest is the length scale of the surface fluctuations and how they
behave dynamically as more material is added.''\cite{ew} And thus was
born the Edwards-Wilkinson (EW) model for surface growth - a solvable
linear model at the heart of our current understanding of numerous
growth processes.  A relevant nonlinear term, added to this by
Kardar-Parisi-Zhang (KPZ)\cite{kpz,halpin,barbasi}, brought to light
the nuances of growth phenomena to the extent that the KPZ equation
very soon became a paradigm, in particular for dynamic phase
transitions. The applicability of the KPZ equation seems to encompass
length scales from atomic level to macroscopic phenomena of every day
life, but still a specter is haunting the field : why is the KPZ
behavior not observed\cite{halpin}?
  
Many of the experimental situations however involve complex processes
which beg to go beyond the idealization as pouring of noninteracting
particles.  This is especially true if medium or fluctuation induced
interactions interfere with the process as for example in the several
recently studied systems involving proteins, colloids or latex
particles \cite{lei,feder,wojt,ramsden}, or in sedimentation.  The
major interaction one has to reckon with, as detailed numerical
computations suggest\cite {pagona1,pagona2}, is the {\it long ranged}
hydrodynamic interaction. Are such long range interactions relevant
for the roughness of the surface?  This question, the absence of a
formalism to handle such interactions in the growth process, and the
elusiveness of the KPZ behavior, led us to propose a simple
phenomenological model by focusing on the long range nature of the
extra force.

We like to develop a Langevin equation type description where long
range aspects can be simulated by a force at each point of the growing
surface exerted by the particles away from it - a hint to go beyond a
strict local description.  In the linear EW model, the growth is along
the global normal to the surface without any overhang. The height
$h({\bf r}, t)$ at point ${\bf r}$ and time $t$ satisfies the
diffusion equation with an additional noise term.  If, instead of the
global, the local normal is favored, the KPZ $(\nabla h)^2$ term is
needed \cite{kpz}.  This nonlinear term describes the lateral growth
at a point as can be seen from the height
profile\cite{halpin,barbasi}. We now extend this physical
interpretation and take the gradient (or its magnitude) as a measure
of the local density of deposited particles.  The long range effect is
now incorporated by coupling these gradients at two different
points. Based on this intuitive picture, the equation we propose is
the following:
\begin{eqnarray}
&& \frac{\partial h({\bf r},t)}{\partial t} = \kappa \nabla^2 h({\bf
    r},t) + \eta({\bf r},t)+ \nonumber \\
&&\frac{1}{2}\ \int d{\bf r'} \ {\vartheta({\bf r'})} \nabla
  h({\bf r + r'},t)\cdot \nabla h({\bf r - r'},t),
  \label{mblong}
\end{eqnarray}
where $\kappa$ is the diffusion constant for the particles on the
surface, and $\eta$ is a random space time dependent white noise of
zero mean and $\langle \eta({\bf r}, t) \eta({\bf r}', t')\rangle = 2
\Delta \delta({\bf r}-{\bf r}')\delta( t-t')$.  The kernel
$\vartheta(r)$ is of long range and, in principle, connected to the
underlying interactions\cite{comm1}. So, we take $\vartheta({\bf r})$
to have a short range part $\lambda_0 \delta({\bf r})$ and a long
range part $\sim r^{\rho-d}$, or more precisely, in Fourier space,
$\vartheta({\bf k}) =\lambda_0 + \lambda_{\rho} k^{-\rho}$.
Eq. \ref{mblong} then smoothly\cite{krugmeakin} goes over to the KPZ
equation for $\lambda_{\rho}=0$.  We show that this leading term
introduced is sufficient to yield a new fixed point with continuously
varying exponents, and different phase transitions not found in the
KPZ problem. The connection with experiments is discussed near the end
of this paper.
 
A central quantity of interest in growth problems is the scaling
behavior of fluctuation of the height, $\langle \mid h({\bf
r},t)-h({\bf 0},0)\mid^2\rangle$ which on a large length and time
scale has a scaling form $ \mid {\bf r}\mid ^{2\chi}{\cal F}({\mid
t\mid}/{\mid {\bf r}\mid ^z})$.  Here $\chi$ is the roughness exponent
of the growing surface and $z$ is the dynamic exponent.  These two
exponents define the universality classes of roughening.

At $d=1$, for the local growth (i.e. KPZ) equation a disorder
dominated rough phase is found for all $\lambda_0$ by several exact
treatments\cite{kardar,huse} providing $\chi=1/2$ and $z=3/2$.  The
nonlinearity is marginally relevant at $d=2$ and for $d>2$ there is a
phase transition from a strong disorder dominated phase ($\chi + z =
2$ for all $d$) to a weak coupling phase where nonlinearity is
irrelevant, i.e.  a flat phase with $z=2$. The perturbation theory is
inadequate for the strong coupling phase at $d\ge 2$ due to the lack
of a perturbative fixed point\cite{frey}. Numerical
simulations\cite{forrest} predict $z\approx 1.6$ at $d=2$. The phase
transition is, however, under control, with $z=2$ $\forall d>
2$\cite{tang,doty}, with a rather complicated critical
behavior.\cite{sm1,sm2}.

A simple scaling analysis indicates that both $\lambda_\rho$ and
$\lambda_0$ are relevant for $d<2$ at the Gaussian fixed point (EW)
where one expects, $\chi=(2-d)/2$ and $z=2$. This follows from the
scale invariance of Eq.(\ref{mblong}) under the transformation
$r\rightarrow br$, $t\rightarrow b^z t$, $h\rightarrow b^\chi h$,
whence $\kappa \rightarrow b^{z-2}\kappa$, $\Delta \rightarrow
b^{z-d-2\chi} \Delta $, $\lambda_0\rightarrow b^{\chi+z-2} \lambda_0$
and $\lambda_{\rho}\rightarrow b^{z+\chi+\rho-2}\lambda_\rho$.  Also
for any nonzero $\lambda_{\rho}$ with $\rho>0$ the local KPZ theory
($\lambda_\rho=0,$ and $\chi +z=2$) is ``unstable'' under
renormalization and a non-KPZ behavior is expected.  For $2<d<2+2\rho$
only $\lambda_\rho$ is relevant at the EW fixed point.  In the
following we adopt a dynamic renormalization group (RG )procedure. 
Our results show a new
stable fixed point at $d=1$, for any $\rho >0$.  Another interesting
consequence of this nonlocality is the possibility of a stable fixed
point at $d=2$ for a certain range of $\rho$. The marginal relevance
of nonlinearity in the original KPZ theory is destroyed.

The renormalization procedure is most succinctly described through the
Fourier modes momentum ${\bf q}$ and frequency ${\bf \omega}$ in terms
of which Eq. (\ref{mblong}) becomes
\begin{eqnarray} 
h({\bf q},\omega)=G_0({\bf q},\omega)[\eta({\bf q},\omega)-
(1/2) (2\pi)^{-d-1}    \times \nonumber\\
\int  d{\bf q}'d{\omega}' \vartheta(2{\bf q}')\
{\bf {q_+}} \cdot {\bf {q_{-}}}\  
h({\bf {q_+}},\omega_+) h({\bf {q_-}},\omega_-)]&&\label{effG},
\end{eqnarray}
\noindent where, symbolically, $X_{\pm}= X/2 \pm X'$ with $X={\bf q}$
or $\omega$.  Here $G_0({\bf q},\omega)$ $=1/(\kappa q^2-i\omega)$
represents the bare propagator or the Green function for the diffusion
equation.  We follow the usual iterative perturbation scheme where $h$
in the RHS of Eq.  (\ref{effG}) is replaced by Eq. (\ref{effG}) itself
upto $O(\vartheta^2)$. A convenient diagrammatic representation can be
set up from this scheme and the renormalization of the various
parameters can be obtained from appropriate vertex functions. We skip
the details as they are very similar to Ref.  \cite{medina}. In the
subsequent renormalization procedure, we integrate out small length
scale fluctuations over a momentum shell $\Lambda e^{-l}\le q'\le
\Lambda$ to obtain the effective parameters for a similar equation but
with a smaller cutoff $\Lambda e^{-l}$, where $\Lambda$ (set to 1) is
related to the microscopic cutoff.  A subsequent rescaling then
restores the cut off to $\Lambda$.

The effective propagator $G({\bf q},\omega)\equiv h({\bf
q},\omega)/\eta({\bf q},\omega)$, gives the renormalization of tension
$\kappa$. The effective noise, obtained from $\langle h^*({\bf
q},\omega)h({\bf q},\omega)\rangle=2{\tilde{\Delta}} G({\bf q},\omega)
G(-{\bf q},-\omega)$, gives the renormalization of the disorder.  Next
we look for the terms contributing to the effective nonlinearity.
Note that the RG transformation, being analytic in nature, cannot
generate a singular term to renormalize $\lambda_{\rho}$ for
$\rho>-2$.  In fact there is no renormalization of $\lambda_0$ either.
A contribution to $\lambda_0$ could come from terms of
$O(\Delta\vartheta^3)$, and a straightforward calculation\cite{medina}
shows that such terms do cancel each other\cite{comm4}.

Following the above procedure we arrive at the flow equations for
$\kappa$ and $\Delta$ as
\begin{eqnarray}
&&\frac{d\kappa}{dl}= \kappa \left [ z- \left (2 +\frac{\Delta
K_d}{\kappa^3}\ \vartheta(2)\vartheta(1)\ \frac{(d-2)+ 3 f(1)}{4d}\right )
\right]\label{kappa}\\
&&\frac{d\Delta}{dl}=(z-d-2\chi)\Delta+\frac{\Delta^2K_d}{4\kappa^3}
\vartheta(2)^2,
\end{eqnarray}
where $f(a) = \partial \ln\vartheta(k)/\partial \ln k\mid_{k=a}$, the
(effective) exponent of $\vartheta(k)$ and $K_d=S_d/(2\pi)^d$, $S_d$
being the surface area of a $d-$dimensional unit sphere.  The flow
equations for $\lambda_0$ and $\lambda_{\rho}$, having contribution
only from the rescaling, are
\label{dlam}
$d\lambda_x/dl =(\chi+z-2 + x)\lambda_x,\quad (x=0\  {\rm or}\  \rho)$.
The two parameters $\chi$ and $z$ are chosen to keep  $\kappa$ and one
of $\lambda_x$ invariant.
 
In terms of $U_x^2={\Delta \lambda_x^2K_d}/{\kappa^{3}}$, and
$R=U_0/U_{\rho}$, with the choice $\chi+z=2$ (or $2 -\rho$) and $z$
equal to the expression inside the big round bracket of Eq. \ref{kappa},
the flow equations can be combined into two as
\begin{eqnarray}
&&\frac{dU_0}{dl} =\frac{(2-d)}{2}U_0+ \frac{2d-3}{4d} U_0^3 +
\frac{U_0U_{\rho}}{8d}[c_0 U_0+  c_1 U_{\rho}],
\label{flow1}
\end{eqnarray}
and $dR/dl = - \rho R$, where $c_0$ = $(5d -6)(1 + 2^{-\rho}) - 2d -
9\rho$, and $c_1=[(3 + 2^{-\rho})d - 6 -9\rho]2^{-\rho}$.  The
equation for $R$ rules out the existence of any off axis fixed point
in the $U_0$ and $U_{\rho}$ parameter space ( except for $\rho=0$,
when there is a trivial marginal fixed line.)

There are only two sets of axial fixed points, SR$\equiv \{
{U_0^{*^2}=2d(d-2)}/{(2d-3)},U_\rho^{*^2}=0\}$, with $\chi + z = 2$,
and LR $\equiv \{U_0^{*^2}=0,\ U_\rho^{*^2}={4d(d-2-2\rho)/c_1}\}$
with $\chi + z = 2 - \rho$. The first set (SR), with $\lambda_\rho
=0$, corresponds to the known KPZ fixed point, whose properties have
already been mentioned.  However we see a relevant perturbation
$U_{\rho}$ which grows at this fixed point.  The stable fixed point
for $d<2+2\rho,$ with $\rho >0$ is LR, except for the region bounded
by $d={(9\rho+6)}/{(2^{-\rho}+3)}$ and $d=2+2\rho$. This excluded
region is, like the KPZ case, an artifact of one loop
renormalization\cite{frey}.  At the new fixed point LR
\begin{equation}
z= 2 +\Phi,\ {\rm and}\ \chi = -\rho - \Phi,
\label{dynam} 
\end{equation} 
where $\Phi=(d-2-2\rho)(d-2-3\rho)/[{d (2^{-\rho}+3 ) - 6-9\rho)}]$
This fixed point admits $z<1$ (not unexpected for long range cases)
but by virtue of the relation $\chi +z = 2 - \rho$, $\chi$ need not be
greater than 1, a requirement for ignoring higher order terms in
Eq. \ref{mblong}.  At $d=2$, the marginal relevance of $U_0$ of the
KPZ theory is lost and there is a stable fixed point (LR) for
$\rho>0.194$.

\vbox{
\psfig{file=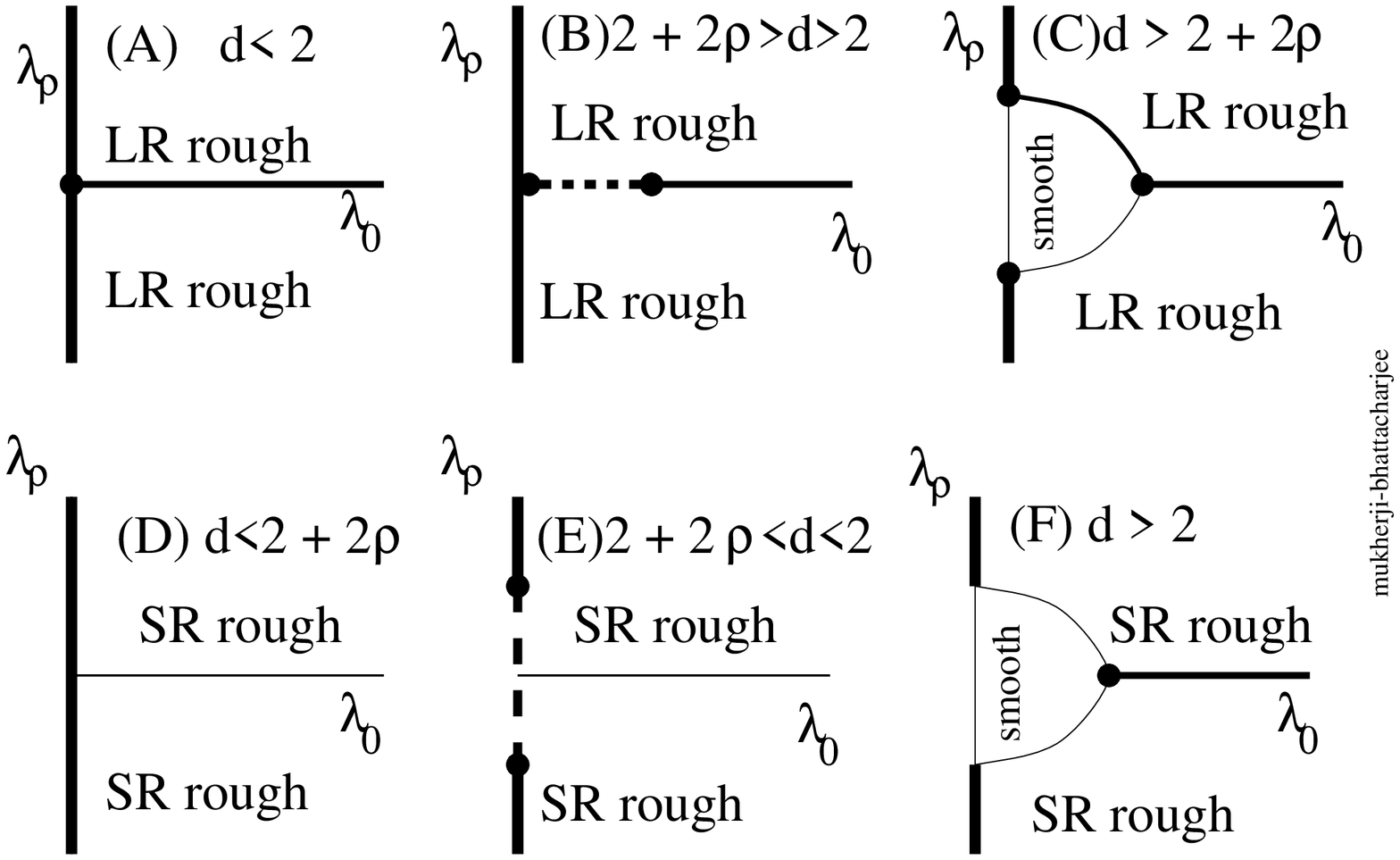,width=2.5in,angle=0}
\begin{figure}
\narrowtext
\caption{ $\lambda_{\rho}\ vs \ \lambda_0$ phase diagram. (A)-(C)
correspond to $\rho>0$, while (D)-(F) to $\rho<0$. Thick line along the
y-axis represents the LR phase, while the medium thick line along the
x-axis SR phase.  In (D) and (E) the phase is SR (KPZ) type for all
$\lambda_\rho\neq 0$. The dashed line in (B) and (E) represents a
smooth phase, which extends over a region in (C) and (F).}
\end{figure}
}

To discuss the surface morphology and the phase
transitions\cite{correl}, we consider different values of $d$ and
$\rho > -2$.  See fig. 1.  Also note that the invariance of
Eq. \ref{mblong} under $h\rightarrow-h$ and $\lambda
\rightarrow-\lambda$, is respected by the fixed point equations.
Since the nonlinear term is like a force, the change in sign of
$\lambda$ corresponds to a ``push''-``pull'' change or a growing to a
receding surface case. We therefore consider both positive and
negative values of $\lambda_{\rho}$, and without any loss in
generality, take $\lambda_0 \geq 0$.

Case I: $d<\min (2,2+2\rho)$: For $\rho>0$, the stable fixed point, if
exists, is LR with the dynamic exponent given by Eq. \ref{dynam}.
Even if it does not exist in this one loop approach, still, from the
flow, the phase is the ``strong'' disorder type. We call this an LR
phase to distinguish it from the SR or KPZ phase.  It is possible to
have a transition between two identical LR rough phases (push-pull).
The critical behavior is EW type if there is strictly no short range
nonlinearity else it is KPZ (SR) type.  See Fig 1A.  In contrast, for
$\rho<0$ (Fig 1D), the LR is irrelevant and the surface behavior is
always SR (KPZ) type except for $\lambda_0 =0$, when it is an LR
phase. There is no phase transition for $\lambda_0 \neq 0$.

Case II: $\min (2,2+2\rho) < d<\max (2,2+2\rho)$: The phases are LR or
SR depending on the sign of $\rho$ (Fig 1B, 1E). For $\rho>0$, the
critical behavior depends on the strength of the SR nonlinearity,
$\lambda_0$.  For small $\lambda_0 <\lambda_{0c}$, the critical
surface is a smooth one while for $\lambda_0 > \lambda_{0c}$ it is
KPZ. There is no transition if $\rho<0$ and $\lambda_0 \neq 0$.
However, for $\lambda_0 =0$, there is a LR rough to smooth transition
for $\rho<0$. See Fig. 1E.

Case III: $d>\max (2,2 + 2\rho)$: For $\rho>0$, the LR fixed-point is
unstable.  A small nonlinearity dies down yielding a smooth surface
while large nonlinearity will produce an LR rough phase.
Unfortunately the absence of a fixed point forbids any prediction of
the behavior of the LR phase.  The unstable LR fixed-point controls
the transition between the rough and the smooth surface with a dynamic
exponent $ z_c = 2 + c \rho \epsilon + O(\epsilon^2)$ where $\epsilon
= d-2 -2\rho$ and $c=-2^{\rho}/[(2 + 2\rho)(1+3\ 2^{\rho} ) - 3\
2^{\rho}(2 + 3 \rho)]$. This is in striking contrast with the believed
to be exact result of $z_c =2$ for the KPZ case. The phase diagram is
shown in Fig 1C.  For $\rho<0$, there is a phase transition between a
SR rough and a smooth phase only if $\lambda_0$ is less than a
critical value as shown in Fig. 1F.

Experiments on colloids\cite{lei} have yielded a value of $\chi=.71$
which is also the value obtained from paper burning
experiments\cite{paper}.  These are taken as the exponent for a driven
surface (line in the $d=1$ example).  For the colloid problem,
hydrodynamic interaction (HI) is important while in the paper burning
experiment, it is possible to have a long range interaction through
the microstructure of the paper.  With this $\chi$ Eq. \ref{dynam} at
the LR fixed point in $d=1$ gives $\rho=-.12$.  At this point, it is
difficult to conclude if this is the transient exponent seen,
eventually going over to the KPZ value on large length and time scales
(Fig. 1D), or it is a true $\lambda_0 =0$ case.  Other cases where HI
is known to play a role, namely the deposition of latex particles or
proteins, the experiments have not been done for roughness of the
growing surface.  We believe such experiments will shed new light on
growth phenomena.

For the KPZ problem, it is known that anisotropy of the substrate can
lead to an overall irrelevance of the nonlinearity in two
dimensions\cite{wolf}.  To see if anisotropy can have a major effect
in the long range case, we now consider a variation of the problem
where the long range interaction has different amplitudes in different
directions.  Restricting ourselves to $d=2$, we take
\begin{eqnarray}
&&\frac{\partial h({\bf r},t)}{\partial t} = \kappa_\parallel
\partial_\parallel^2 h({\bf r},t)+\kappa_\perp
\partial_\perp^2 h({\bf r},t)+
\eta({\bf r},t)+\nonumber\\
&& \sum_{\Psi=\parallel,\perp}
\int d{\bf r'} \ \frac{1}{2} \vartheta_\Psi({\bf r'}) \partial
_\Psi h({\bf r+r'},t) \partial_\Psi h({\bf r-r'},t). 
\end{eqnarray}
as the anisotropic version of Eq. \ref{mblong}.  In the isotropic
case, $r_\lambda\equiv\vartheta_\parallel({\bf
r})/\vartheta_\perp({\bf r})=1$ and
$r_\kappa\equiv\kappa_\parallel/\kappa_\perp=1$ reproduce eqn
\ref{mblong}. For simplicity let us concentrate only on a case of
anisotropy in the long range part, with $\lambda_0 =0$,
$\lambda_{\perp}({\bf q})=\lambda_{\perp \rho} q^{-\rho}$ and
$\lambda_{\parallel}({\bf q})=\lambda_{\parallel \rho} q^{-\rho}$.  An
anisotropic scaling of the surface $x_\perp\rightarrow e^l x_\perp$
and $x_\parallel\rightarrow e^{l\zeta}x_\parallel $ lead to $r_\lambda
\rightarrow e^{2(1-\zeta)l} r_\lambda $.  For nonzero $r_\lambda$ the
scale invariance consequently restricts $\zeta=1$. For $r_\lambda=0$,
this constraint cannot be imposed and the analytical tractability is
lost. To avoid this complexity here, we take $r_\lambda \not= 0$ and
$\zeta =1$.
 
The RG procedure follows as before, only a new flow equation for
$r_\lambda$ is required. The recursion relations  are
\end{multicols}
\widetext
\rule{5cm}{.5mm}
\begin{eqnarray}
\frac{d\kappa_\perp}{dl}=(z-2)\kappa_\perp+\frac{g_\perp
  \kappa_\perp}{16
r_\kappa^{1/2}2^{\rho}}\left(1-\frac{r_\lambda}{r_\kappa}\right)+\frac{3\rho
g_\perp   \kappa_\perp}{4
r_\kappa^{1/2}2^{\rho}}\frac{(r_\kappa+r_\lambda+2{\sqrt
r_\kappa})}{(1+{\sqrt r_\kappa})^2}\label{anikap}\\  
\frac{dr_\kappa}{dl}=-\frac{g_\perp r_\kappa^{1/2}}{16 \times 2^{\rho}}
\left(1-\frac{r_\lambda^2}{r_\kappa^2}\right)- 
\frac{3\rho g_\perp 2^{-\rho}}{4 r_\kappa^{1/2}(1+{\sqrt r_\kappa})^2}
\left(2r_\kappa^{3/2}+r_\kappa^2+r_\lambda
r_\kappa-r_\lambda-\frac{r_\lambda^2}{r_\kappa}-\frac{2r_\lambda^2}{{\sqrt
r_\kappa}}\right)\label{rkap}\\
 \frac{dg_\perp}{dl}=2\rho g_\perp+
\frac{g_\perp^2 2^{-\rho}}{16 r_\kappa^{1/2}} 
\left[\frac{(3r_\lambda^2+3r_\kappa^2+2 r_\lambda
r_\kappa)}{r_\kappa^{2}2^{1+\rho}}-   
3 \left(1-\frac{r_\lambda}{r_\kappa}\right)-
36\rho \frac{r_\kappa+r_\lambda+2{\sqrt r_\kappa}}{(1+{\sqrt
r_\kappa})^2}\right],   
\end{eqnarray}
{}\hfill\rule{5cm}{.5mm}
\begin{multicols}{2}
\noindent where $g_{\perp} = \lambda_{\perp}^2\Delta
K_2\kappa_{\perp}^{-3}$. For $r_\lambda >0$, Eq. \ref{rkap} has a
fixed point with $r_\kappa^* >0$, which is a continuation of the
isotropic fixed point $r_\lambda=r_\kappa=1$ for $\rho=0$. The
important fixed point for us is $r_\kappa^*=-r_\lambda$, which is
physical, from the stability requirement of the surface, only if
$r_\lambda$ is negative. We consider only this anisotropic case
here. The flow equation for $g_\perp$ now allows a fixed point, unlike
the isotropic case discussed earlier. For small $\rho$, the
anisotropic fixed point is at $g_\perp^*\approx 8 \rho \sqrt{\mid
r_\lambda\mid}$, with $z=2 - \rho/2 + O(\rho^2)$ from
Eq. \ref{anikap}.

The effect of different signs of $\lambda_\rho$ is to have opposing
(push/pull) effects in the two orthogonal directions.  In the KPZ
case, they cancel each other producing a EW surface\cite{wolf}.  In
the LR case, we predict a new type of rough surface with $z<2$ for
$\rho>0$ though on the whole it may be flat but singular with
$\chi<0$. Surprisingly, this case seems to be better controlled in the
RG approach than the isotropic case.  This, in turn, calls for further
studies of the $r_\lambda=0$ situation entailing anisotropic scaling
of space ($\zeta \neq 1$).

In summary, we have proposed a simple phenomenological model,
Eq. \ref{mblong} that incorporates, as a minimal model, long range
interactions in growth problems.  We have shown that {\it any}
interaction decaying slower than $1/r^d$ makes the KPZ or the short
range nonlinear case unstable, and asymptotically the surface will
have different roughness with exponents depending on the power law of
the interaction.  The critical behavior in going from a growing to a
receding surface can be of various types depending on the
dimensionality and the strength of the interaction, as shown in Fig 1.
Power laws decaying faster than $1/r^d$ are suppressed by any local or
short range nonlinearity yielding a KPZ like roughness, but when
alone, it can produce a still rougher surface.

 SM acknowledges support from SFB341.  SMB thanks
T. Nattermann for hospitality at K\"oln. 

\end{multicols}


\vskip -.5cm
\begin{references}
\vskip -.5cm
\bibitem[*]{eml1}email: sutapa@thp.uni-koeln.de; sb@iop.ren.nic.in
\bibitem{ew}S. F. Edwards and D. R. Wilkinson, Proc. Roy. Soc. {\bf A
    381}, 17 (1982).
\bibitem{kpz} M. Kardar, G. Parisi and Y. C. Zhang,
  Phys. Rev. Lett. {\bf 57}, 1810 (1986).
\bibitem{halpin} T. Halpin-Healy and Y. -C. Zhang, Phys. Rep. {\bf
    254}, 215 (1995). J. Krug, Adv. Phys. {\bf 46}, 139 (1997).
\bibitem{barbasi} A. L. Barb\'asi and H. E. Stanley, {\it Fractal
concepts in surface growth} (Cambridge University Press, NY, 1995).
\bibitem{lei} Xin-Ya Lei et al, Phys. Rev. E {\bf 54}, 5298 (1996).
\bibitem{feder} J. Feder and I. Giaever, J. Colloid Interface
Sci. {\bf 78}, 144 (1980).
\bibitem{wojt} P. Wojtaszczyk et al, J. Chem. Phys., {\bf 99}, 7198
(1993); J. Chem. Phys. {\bf 103}, 8285 (1995).
\bibitem{ramsden} J. J. Ramsden, Phys. Rev. Letts. {\bf 71}, 295
(1993).
\bibitem{pagona1} I. Pagonabarraga and J. M. Rubi, Phys. Rev. Letts.
{\bf 73}, 114 (1994).
\bibitem{pagona2} I. Pagonabarraga et al, J. Chem. Phys., {\bf 105},
7815 (1996).
\bibitem{comm1} The nonlinear term chosen does not have a unique sign.
It can easily be cured, e.g., by taking the absolute value, but
at a cost of complexity. The differnece between the chosen form and a
direct coupling of ${\bf r}$ and ${\bf r'}$ is irrelevant for
universal behavior.
Also not considered in the 
paper are quenched noise and the driven case. 
\bibitem{krugmeakin} J. Krug and P. Meakin, Phys. Rev. Letts. {\bf
    66}, 703 (1991) consider a case where nonlocality is introduced in
  the linear model.  This however does not go continuously to the KPZ
  case - a feature manifest in Eq. \ref{mblong}. Also to be noted is a
different proposal, K. B. Lauritsen, Phys. Rev. E {\bf 52}, R1261
(1995). 
\bibitem{kardar} M. Kardar, Nucl. Phys. B {\bf 290}, 582 (1987)
\bibitem{huse} D. A. Huse etal,  Phys. Rev. Lett. {\bf 54}, 2708 (1985)
\bibitem{frey}There is a controversy over the two loop RG result and
  the fixed point in $d=2$. T. Sun and M. Plischke, Phys. Rev. E {\bf
    49}, 5046 (1994); T. Sun, ibid {\bf 51}, 6316 (1995); E. Frey and
  U. C. Tauber, Phys. Rev. E {\bf 50}, 1024 (1994); {\bf 51}, 6316
  (1995); E. V. Theodorovich, JETP {\bf 82}, 268 (1996);  K. J. Wiese,
  cond-mat/9706009 reanalyzed all of these two loop calculations.
\bibitem{forrest} B. M. Forrest and L. -H. Tang, Phys. Rev. Lett. {\bf
    64}, 1405 (1990)
\bibitem{tang} L. H. Tang etal, Phys. Rev. Lett. {\bf 65}, 2422 (1990).
\bibitem{doty} C. Doty and J. M. Kosterlitz, Phys. Rev. Lett. {\bf
    69}, 1979 (1992)
\bibitem{sm1} S. Mukherji, Phys. Rev E {\bf 50} R2407 (1994).
\bibitem{sm2} S. Mukherji and S. M. Bhattacharjee, Phys. Rev. B {\bf
53}, R6002 (1996).
\bibitem{medina} E. Medina {\it et al}., Phys. Rev. A {\bf 39}, 3053
  (1989)
\bibitem{comm4} That there is no renormalization of $\lambda$'s can be
seen easily in a Martin-Siggia-Rose type field theoretic formulation.
(See, e.g., Ref \cite{frey}, S. Mukherji Phys. Rev. {\bf E55}, 6459
(1997)). Since the $\lambda$ vertex involves only equal time fields,
and for  causality the response propagator is nonlocal
in time, the triangular diagrams ($O(\lambda^3)$) that could
have renormalized $\lambda$, vanish identically.
\bibitem{correl} It is important to recognize that the results are even
  qualitatively different from the case of long range correlated noise
  in the KPZ equation\cite{medina}, mainly because the long range part
  of Eq. \ref{mblong} does not generate any short range piece under
  renormalization. 
\bibitem{paper} J. Zhang etal, Physica A {\bf 189} 383 (1992)
\bibitem{wolf} D. E. Wolf, Phys. Rev. Lett.  {\bf 67} 1783 (1991).
\end{references}
\end{document}